# DARK MATTER FORMATION AT SCHWARZ SCALES: PRIMORDIAL FOG PARTICLES AND WIMP SUPERHALOS

CARL H. GIBSON

University of California at San Diego, Departments of Applied Mechanics
and Engineering Sciences and Scripps Institution of Oceanography,
La Jolla, CA 92093-0411, USA, cgibson@ucsd.edu

Dark matter appears in two forms as a consequence of the fluid mechanics of self-gravitational condensation. Condensation occurs primarily on non-acoustic nuclei rather than on the acoustic nuclei of the Jeans (1902) criterion, leading to a very different scenario for structure formation. Viscous forces at $10^{12}$ s (30,000 y) after the big bang permit decelerations of $10^{47}$ kg protosupercluster plasma masses, and $10^{42}$ kg protogalaxy masses at $10^{13}$ s (300,000 y). Then gas formed, and all the baryonic universe became a "primordial fog" of $10^{23}$ kg particles at the viscous Schwarz scale: 100% dark matter. Some of these H-He objects have collected to form stars, but most persist as dark dwarfs in galaxy halos. They manifest themselves in quasar microlensing observations as "rogue planets", Schild (1996), "dark galaxies", Hawkins (1996), and as comets "brought out of cold storage", O'Dell and Handron (1996). Non-baryonic WIMP fluids are superviscous, with large viscous Schwarz scales, and condense slowly to form most of the dark matter of galaxy superclusterhalos and clusterhalos.

## 1 Introduction

Most discussions of structure formation in the universe rely exclusively on the classical analysis of the English astrophysicist Sir James Jeans[1,2]. Jeans considered ideal, inviscid, irrotational flows, leading to condensation on density nuclei propagating with the speed of sound. Refinements of the Jeans theory[3] preserve the linear perturbation flavor of his analysis and change his central conclusion only slightly; that is, that the minimum size of self-gravitational condensation is at a length scale $L_J = V_s/(\rho G)^{1/2}$, where $V_s$ is the speed of sound, $\rho$ is the density of the gas, and G is Newton's gravitational constant $6.67 \times 10^{-11}$ m$^3$ kg$^{-1}$ s$^{-2}$. Jeans' incorrect assumption that his acoustic criterion is an "adequate ... explanation of the creation of ... astronomical bodies" (p345) has been compounded by an erroneous concept that the internal pressure of the gas resists gravitational collapse[4]. It is suggested here that condensation on non-acoustic nuclei at Schwarz scales is generally the dominant process of gravitational structure formation, and is not resisted by internal pressure[5] (pressure forces require pressure gradients). A physical model corresponding to the "pressure" derivation of the Jeans length $L_J$ is shown on the left in Figure 1, compared to a physical model and derivation of the viscous Schwarz length $L_{SV}$ on the right.

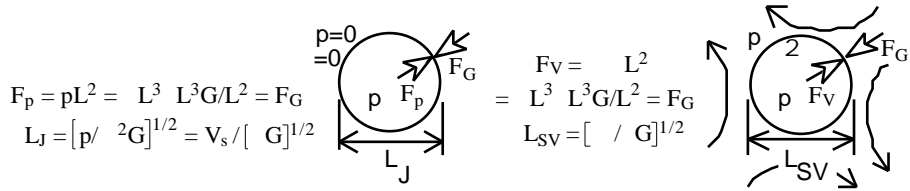

Figure 1: Pressure model for Jeans scale (left) assumes the condensate is in an infinite vacuum. Viscous Schwarz scale model (right) for a non-acoustic density perturbation.

The Jeans mass $M_J$ vastly overestimates the minimum masses of condensation by gravity in the early stages of the universe when most structures formed. Because $V_s$ c during the plasma epoch, no plasma condensation is permitted because the Jeans scale $L_J$ is always larger than the Hubble scale $L_H = ct$, where c is the speed of light and t is the time since the big bang (information about density fluctuations on scales larger than $L_H$ cannot be transferred in times less than t). At neutralization an $M_J$ of $10^{36}$ kg is found[4].

Since condensation on non-acoustic nuclei is not inhibited by the "Jeans pressure" $p_J = V_s^2$ it begins in the plasma epoch as soon as the viscous Schwarz radius $L_{SV} = (\ /\ G)^{1/2}$ decreases to values less than $L_H$, where is the rate of strain of the fluid and is the kinematic viscosity[5]. Kinematic viscosities $L_p x V_p$ of cosmic fluids during the initial stages of the hot big bang model are enormous compared to terrestrial values because distances between particle collisions $L_p$ and particle speeds $V_p$ are much larger. Therefore condensation occurs at $L_{SV}$ scales rather that at the smaller turbulent Schwarz scales $L_{ST} = (\ )^{1/2}/(\ G)^{3/4}$, where is the viscous dissipation rate, because Reynolds numbers of the flows at the condensation scale are subcritical[5]. Cosmic Background Experiment (COBE) satellite data give small shear and vorticity[6] estimates that confirm the assumption of weak or no turbulence.

## 2 Structure Formation in the Universe

The first self-gravitational condensation occurs when $L_{SV} = L_H$ as $L_{SV}$ decreases and $L_H$ increases with time. The time can be estimated by equating the mass of the first condensate $M_{SC}$ to $L_H^3$, where (t) is the critical density of the universe as a function of time computed from Einstein's general field equations[7] (p540, table 15.4), $L_H(t) = ct$, and $M_{SC}$ is the mass of galaxy superclusters, which should be about $10^{47}$ kg based on observations that the largest supervoid scale is $2 \times 10^{24}$ m. Taking the scale of the visible universe to be $2 \times 10^{26}$ m and the mass to be $10^{53}$ kg gives $M_{SC} = 10^{47}$ kg. A plot of (t)$c^3 t^3/M_{SC}$ versus t crosses 1 at t = $10^{12}$ s, which is taken to be the time of the first self-gravitational condensation, 30,000 years after the big bang. Setting $L_H = L_{SV}$ at this time gives  = $10^{28}$ m$^2$ s$^{-1}$ and a barely subcritical Reynolds number, confirming the assumption that the flow is non-turbulent at horizon scales (the largest possible for turbulence). The maximum possible   value $c^2 t$ at this time is about $10^{29}$ m$^2$ s$^{-1}$, an order of magnitude larger than our estimate.

With further cooling the viscous Schwarz scale $L_{SV}$ increases less rapidly than $L_H$, permitting condensation (deceleration of expansion) of smaller and smaller masses to about $10^{42}$ kg before plasma neutralization[5]. This is the mass of a galaxy. Because a preferred condensation axis always exists in a condensing blob of slightly nonspherical gas, the expected topology of the protosupercluster to protogalaxy condensates formed in the period $10^{12}$ to $10^{13}$ s is that of a Zel'dovich nested foam. Figure 2 shows the large to small mass sequence of condensation events during the plasma epoch.

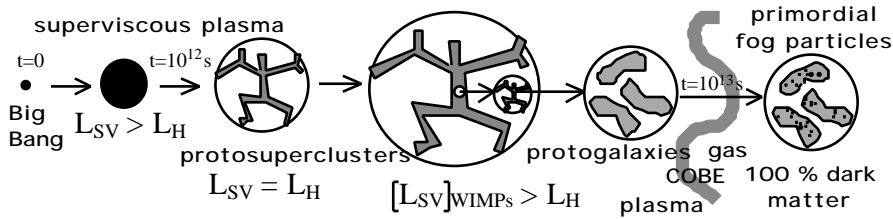

Figure 2: Top-down self-gravitational condensation during the plasma epoch.

Because photons decouple from matter at the plasma-gas transition, the viscosity sharply decreases. The condensation mass is that of a "primordial fog particle" (PFP) $M_{PFP} = L_{SV}^3$  = $(\mu /G)^{3/2}/^2$ = $(4.4 \times 10^{-5} \times 10^{13}/6.67 \times 10^{-11})^{3/2}/(10^{-17})^2$ = $1.7 \times 10^{23}$ kg, where μ is estimated from standard tables, the rate of strain   is taken to be 1/t, and the density   is taken to be that of a globular cluster. Increasing   or decreasing   cause increases in $M_{PFP}$. Decreasing   also increases the condensation time, so the first condensation after photon decoupling should be such moon-mass objects (mPFPs) within $10^{36}$ kg Jeans mass protoglobularcluster (PGC) gas droplets. Any lower density protogalaxies should condense later to larger mass PFPs, perhaps up to Jupiter mass (JPFPs). The first star formation requires an accretion of PFPs. Although more mPFPs are needed to make a star, their collision probability is higher than the larger JPFPs in less dense protogalaxies. Thus, galaxies with moon-mass PFPs should form the first stars within PGCs, globular clusters of stars, and globular clusters of clusters. Protogalaxies with Jupiter-mass or larger PFPs might never form stars and remain permanently dark. Figure 3 shows the small to large mass sequence of PFP accretion events expected during the gas epoch, extending to the present. With time, PFPs should become increasingly compact and more weakly collisional.

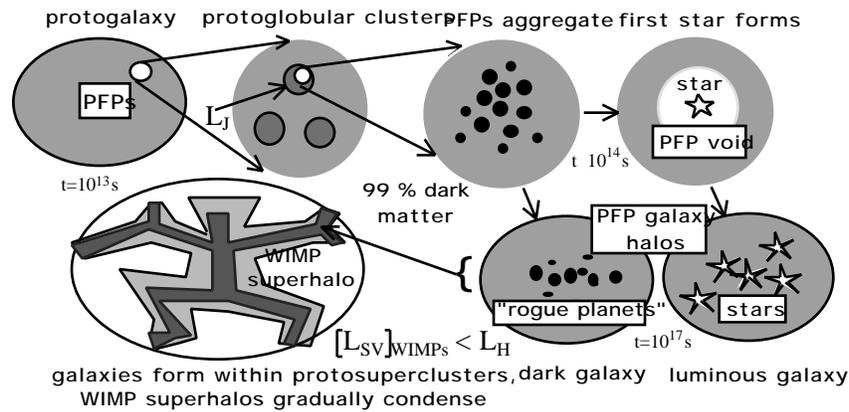

Figure 3: Bottom-up aggregation of PFPs to form stars and other non-dark matter.

## 3  Comparison to Observations

Many recent observations are available for comparison to the present structure formation model. Primordial fog particles predicted above are clearly a candidate for the baryonic dark matter expected to exist in the halos of galaxies, and in dark galaxies where the particles are so large that accretion to stars may never or rarely happen. For example, moon-mass PFPs should initially be separated by distances of only about $10^{14}$ m, but brown dwarf mass PFPs would initially be separated by more than $10^{17}$ m so they have little chance of eventual accretion with other such large PFPs to form stars.

Because PFPs are so small and distant from each other they cannot be observed directly. Estimated angular sizes from Earth are about $10^{-10}$ radians or less, beyond the resolution of even the Hubble Space Telescope (HST). One method of seeing them is provided by nearby dying stars with stellar winds that form planetary nebulae. The nearest planetary nebula to the Earth is the Helix Nebula, and this is full of possible PFPs. O'Dell and Handron[8] estimate over 3500 "cometary globules" with masses of order $10^{25}$ kg, separated from each other by the expected distance of about $10^{14}$ meters. The authors favor a Reyleigh-Taylor instability origin for the objects (which seems highly questionable) but do not rule out the interpretation favored here that they are "highly volatile comets brought out of cold storage" by the powerful winds and radiation from the hot, bright, central star. The suggestion here is that rather than stars of the Milky Way being surrounded by local Oort clouds of comets at $10^{16}$ m, that the Oort cloud of "comets" is continuous between stars, containing Earth-mass PFPs with occasional $10^{16}$ m voids where about $10^6$ PFPs condensed to form each star (Fig. 3).

Another way to detect dark matter objects in galaxy halos is from their microlensing of quasars. The MACHO and EROS star microlensing projects have so far not been able to disprove the existence of $10^{23}$ kg objects, although some doubt has been cast on the existence of the large numbers of $10^{25}$ kg objects inferred from quasar microlensing by Schild[9], and strong doubt for the $10^{27}$ kg objects of Hawkins[10]. In principle, the smaller, brighter, sources of light at greater distances provided by quasars, with greater foreground surface mass density, should be much more productive of robust PFP microlensing events than the nearby star fields employed by MACHO and EROS.

Schild[9] provides a remarkable three year (almost nightly) record of triple image pairs of the brightness of the two quasar image components of the first discoveredgalactic lens Q0957+561 A,B, with occasional images for the 15 years since its discovery, making a 90 year historical record. The records show continuous microlensing activity in both images, with dozens of well resolved 90 day events corresponding to "rogue planet" masses of about $10^{25}$ kg, supporting his conclusion that "these are likely to be the missing mass". Another population of week long events was identified which may either be moon-mass objects in the lens galaxy, or planet mass objects in the Milk Way halo so far undiscovered by the MACHO observations. Refsdal & Stabell[11] suggest that microlensing of quasar Q2237+0305 can also be explained by a population of $10^{25}$ kg mass microlenses, but ask for more accurate lightcurves over longer time spans, such as those given by Schild[9]. Schild[9] presents evidence that an intrinsic 3 year variability period exists in the quasar Q0957+561 A,B images.

Hawkins[10] suggests that most if not all distant quasars are being microlensed by populations of $10^{27}$ kg mass bodies in intervening galaxies (most of which are dark galaxies) sufficient to make up much of the critical density. Evidence for dark galaxies is the observation of Jupiter period microlensing for all observed distant quasars but not for those nearby (z < 0.3). Of the eight known double quasars with image separations of 2-7 seconds, only two have luminous lens galaxies. Six have none, suggesting dark to luminous mass ratios of greater than 1000. The 3 year intrinsic quasar variability period noticed by Schild[9] coincides with the period expected for Jupiter microlensing, suggesting some fraction of the apparent Jupiter microlensing may be intrinsic. Further high frequency sampling of multiple lens quasars for intrinsic variability frequencies should help determine the value of the fraction. Hawkins (this volume, Fig. 2) shows nonintrinsic variability at Jupiter frequencies exists in the double quasar Q2138-431. Microlensing signatures should be identical for the Jupiter mass primordial black holes postulated by Hawkins[10] and the Jupiter mass "primordial fog particles" postulated here.

## 4  Summary

Two forms of dark matter emerge from a self-gravitational condensation theory based on Schwarz viscous and turbulent scale limited condensation on non-acoustic density nuclei.

Baryonic (ordinary) dark matter consists of "primordial fog particles" (PFPs) of H-He primordial matter with masses probably in the range $10^{23}$ to $10^{27}$ kg, presumably now in solid or liquid state and crusted with 14 billion years of accreted dust. Evidence supporting the existence of this primary building material for the structures of the universe is provided by recent quasar microlensing observations of "rogue planets", "dark galaxies", and "Jupiters" with such a high observational frequency that Schild[9] and Hawkins[10] both conclude they are likely to be the missing baryonic dark matter of galactic halos. Cometary globules of the Helix[8] and other planetary nebulae provide further evidence of PFP dark matter. A second form of dark matter predicted by the theory consists of WIMP fluid slowly condensing at large viscous Schwarz scales to form galaxy supercluster halos and galaxy cluster halos on these largest mass objects of the universe.

**Acknowledgments**
The author is grateful for the excellent advice and answers to numerous questions provided by participants at the Sheffield dark matter workshop, particularly Rudy Schild, who was also kind enough to carefully review a preliminary draft of the present paper, and Mike Hawkins who supplied a preprint of his paper.